\documentclass[preprint,aps,amsmath,amssymb]{revtex4-1}

\usepackage{graphicx}
\usepackage{graphicx}

\usepackage{hyperref}

\newcommand{\be}{\begin{equation}}
\newcommand{\ee}{\end{equation}}
\newcommand{\bea}{\begin{eqnarray}}
\newcommand{\eea}{\end{eqnarray}}
\newcommand{\ba}{\begin{array}}
\newcommand{\ea}{\end{array}}

\begin{document}

\title{Hard gluon evolution of nucleon Generalized parton distributions  in the Light-front quark model}
\author{Neetika Sharma }

\affiliation{ 
Indian Institute of Science Education and Research Mohali,\\
S.A.S. Nagar, Mohali-140306, Punjab, India.
}

\begin{abstract}

We  incorporate the  perturbative evolution effects in  the generalized parton distributions (GPDs) calculated in effective light-front quark model for the nucleon.  The perturbative effects enters into formalism through the evolution of GPDs according to the Dokshitzer-Gribov-Lipatov-Altarelli-Parisi-like (DGLAP) equation. We  obtain the evolved GPDs in the momentum space and  transverse impact parameter space. We observe that combining the light front quark model with the  perturbative evolution effects,  give the effective model for studying the phenomenological GPDs.

\end{abstract}
\maketitle


 
\section{Introduction}

Quantum chromodynamics (QCD) is the widely accepted fundamental description of strong interaction in terms of quark and gluon degrees of freedom. It has been proven successful  in explaining the physical phenomena at high-energy range, however, the applicability of QCD to low energies is limited to some extend. Because of color confinement no quarks and gluons have ever been directly observed by any detector in high energy scattering experiments.  QCD factorization  theorem enables us to connect the dynamics of quarks and gluons to physically measured hard scattering cross sections of the known spectrum of hadrons, by systematically  factorizing the physics taking place at different momentum scales \cite{dglap1,dglap2,dglap3,dglap4}.  

Both  the exclusive and inclusive processes, can be factorized  into the perturbative and non perturbative part.  The scattering of the virtual photon off the parton is the short distance part  can be evaluated using the perturbation theory.  The  universal long-distance part is  parametrized in terms of  PDFs, GPDs, or other kinds of form factors and require the knowledge of either non-perturbative methods or a global fit to experimental data. The  initial distributions of quark and gluon are parameterized  as functions of longitudinal momentum of quarks $x$ for a chosen initial scale $\mu^2$ and  then evolved  to numerically  larger values. The independence of  physical observable from the physical scale leads to the DGLAP equation in the perturbative QCD.

The study of GPDs  have been of enormous interest as they contain vital information about the  3-D structure information of the nucleon \cite{gpdrev1,gpdrev2,gpdrev3}. Many models have been proposed theoretically in the recent past to explain the hadronic properties in terms of GPDs \cite{diehl}. Primarily, GPDs are  parametrized in terms of three variables, namely, longitudinal momentum of quark $x$, the invariant momentum transfer $t$, and the skewness parameter $\zeta$, which  gives the fraction of the longitudinal momentum transfer to the nucleon in the process.  The recent experiments are performed at  high-luminosity with large momentum transfer  and  give remarkable precise data for the measurement of  GPDs.   Several experiments, for example, H1 Collaboration \cite{h1}, ZEUS Collaboration \cite{zeus} at HERA collider, HERMES  at DESY \cite{hermes}, have already collected data for the deeply virtual Compton scattering experiments. Experiments presently running at Hall A and B at Jefferson Laboratory \cite{jlab1,jlab2,jlab3}, COMPASS at CERN \cite{compass}, etc., will provide more accurate data in a wider kinematic range. This will significantly help us in advancing  our present understanding of hadron structure.

In a recent  work, Gutsche {\it et al.}  \cite{valery2014,valery2015} have proposed a new phenomenological light-front wave  function (LFWF) for nucleon.  The effective LFWF is derived from light front holography principle of matching the predictions of soft-wall model of AdS with the light-front QCD for electromagnetic form factors (EFFs) of mesons \cite{adsqcd1,adsqcd2,adsqcd3,adsqcd4}.  A new phenomenological light-front quark model (LFQM) has been formulated based on LFWFs considering the nucleon as the quark-scalar diquark bound state, which successfully explain the experimental data for the form factor  of nucleon and their flavor decomposition into up and down quark. 

In this work, we are interested in understanding the effect of perturbative evolution on the phenomenological LFQM \cite{valery2014}  and study the  observable related to the nucleon structure. In particular,  we incorporate the perturbative effects on the  GPDs  calculated in the LFQM and refer them as  ``evolved GPDs''.   It  is necessary to take  into account the hard perturbative evolution  effects to make theory independent of factorization scale.   In addition, we also investigated the  DGPAP evolution of the GPDs in impact parameter space. Impact parameter GPDs provide the  tomographic picture of the distribution of  a quark with momentum fraction $x$ located at a transverse position $b_{\perp}$ from the center of nucleon \cite{impact}.

This paper is structured as follows: In Sec. \ref{formf}, we present the results for nucleon LFWFs  of the quark-scalar diquark model  and present the essential  calculations of valence GPDs in the momentum space.  In the next section \ref{dglap}, we will discuss briefly the way perturbative corrections enter into the formalism giving the DGLAP like evolution of  GPDs.  A detailed comparison of  behaviour  of  GPDs in the transverse impact parameter space is presented in Sec. \ref{impact}.   Summary and conclusions are discussed in Sec. \ref{conc}.


\section{GPDs in the light front quark-scalar diquark model}
\label{formf}

The hard exclusive reactions have found increased attention in the recent past because of new experimental and theoretical developments. Such reactions, for instance, deeply virtual compton scattering where all the kinematical parameters of initial and final particles are measured, contain much more information about the nucleon structure. Generalized parton distributions enter into the factorization theorem for hard exclusive processes and play the role of  long-distance non-perturbative part  in a similar manner as PDFs enter factorization for inclusive DIS processes.  

In this section, we will revisit the essential of   calculations of GPDs in the light front quark model based upon the light front holography principle. The details pertaining to the perturbative evolution effects will be considered in the  next section.  Valence  GPDs  are calculated using  a phenomenological LFWF for the nucleon, which consider the nucleon as bound state of an active quark and a spectator scalar diquark. The LFQM has been able to  successfully produce the EFFs of nucleons including their flavor decompositions consistent with data \cite{valery2014}. 

First, we will recollect the known information about the Dirac $F_1(q^2)$ and Pauli form factors $F_2(q^2)$ for the spin $1/2^+$ particles. In the light-front formalism, it is convenient to identify  the $F_{1,2}(q^2)$ form factors by the helicity conserving and the helicity non-conserving matrix element of the plus component of the electromagnetic current $(J^+)$. 
\bea \left \langle P +q , \uparrow | \frac{J^+ (0)} {2 P^+}| P, \uparrow \right \rangle &=& F_1 (q^2)\,, \\ \left \langle {P +q , \uparrow} | \frac{J^+ (0)} {2 P^+}| P, \downarrow \right \rangle &=& - (q^1 - \iota q^2)\frac{F_2 (q^2) }{ 2 M_N} \,, \eea
where $q^{\mu} = (P'- P)^{\mu}$ is the momentum transferred, $M_N$ is the nucleon mass. Dirac form factors $F_1^{p/n}(0)$ are normalized to  electric charge ($e_{p/n}$) and Pauli form factor $F_2^{p/n}(0)$ to the anomalous magnetic moment ($\kappa_{p/n}$) of the nucleons.

The well known Ji's sum rules that relate the electromagnetic form factors  with the GPDs for unpolarized quarks  \cite{jisum} :
\bea  \label{sumrule1}
F_1^q(q^2) &=& \int_{0}^1 dx\; H^q(x,q^2) \,,\\
F_2^q(q^2) &=& \int_{0}^1 dx\; E^q(x, q^2) \,,
\label{sumrule2}\eea
where we have used the definitions of GPDs with  suppressed skewness $H^q(x,q^2)=H^q(x, 0, q^2) +H^q(-x, 0, q^2);  ~ E^q(x, q^2)=E^q(x, 0, q^2)+E^q(-x, 0, q^2)\,. $ 
The value of GPDs at $(-x)$  for quarks is equal to  GPDs at $(x)$ for antiquarks with a minus sign.   The skewness dependence drops out from the sum rules because of the Lorentz invariance \cite{radyushkin}.  We therefore restrict our study to the case $\zeta=0$ and use the convention $ H^q(x,q^2)$ ($E^q(x,q^2)$) instead of $ H^q(x,\zeta=0, q^2)$ ($E^q(x,\zeta=0, q^2)$.

Considering the proton as the bound state of two particles: a quark and a scalar diquark in the light-front quark model, the spin flip and non-flip GPDs  for the quarks can be written in the light-front representation as  \cite{stan}
\be
H(x, q^2) = 
\int { {\mathrm d}^2 {{\overrightarrow k}_{\perp }} \over 16 \pi^3}
\bigg[  
\psi^{*\uparrow}_{1/2} (x, {\overrightarrow k'}_{\perp}) \, \psi^{\uparrow}_{1/2} (x, {\overrightarrow k}_{\perp})+ 
\psi^{*\uparrow}_{-1/2} (x, {\overrightarrow k'}_{\perp}) \, \psi^{\uparrow}_{-1/2} (x, {\overrightarrow k}_{\perp}) \bigg]\,,
\label{dirac} \ee
\be
E(x,q^2) = {-2 M_N \over q_1 -\iota q_2} \int { {\mathrm d}^2 {{\overrightarrow k}_{\perp }} \over 16 \pi^3} \bigg[  \psi^{*\uparrow}_{1/2} (x, {\overrightarrow k'}_{\perp}) \, \psi^{\downarrow}_{-1/2} (x, {\overrightarrow k}_{\perp})+ 
\psi^{*\uparrow}_{1/2} (x, {\overrightarrow k'}_{\perp}) \, \psi^{\downarrow}_{-1/2} (x, {\overrightarrow k}_{\perp}) \bigg]\,,
\label{pauli}\ee
where $x$ is the fraction of momentum carried by active quark, $t =-Q^2=-q^2_{\perp}$ is the square of momentum transferred, and ${\overrightarrow k'_{\perp}} = {\overrightarrow k_\perp} + (1-x) {\overrightarrow q_\perp}$ is transverse momentum of the  parton.  Also, $\psi^{\lambda_N}_{ \lambda q}(x, k_{\perp})$ is the LFWF describing the interaction of  quark and a scalar diquark to form a nucleon.

We adopt the generic ansatz for the  valence Fock state of the nucleon LFWFs  in the quark-diquark model at an initial scale $\mu_0$=0.3 GeV as proposed in \cite{valery2014}. The explicit form of LFWFs for the spin $1/2$ particles read as 
\bea
\psi^\uparrow_{1\over 2}(x, { k}_{\perp} ) &=& \varphi^1_q(x,\, { k}_{\perp})\,,
\nonumber \\
\psi^\uparrow_{-{1\over 2}}(x, { k}_{\perp} ) &=& - \left( {k^1 + \iota k^2 \over x M_n} \right) \,\varphi^2_q(x,\, { k}_{\perp})\,, \nonumber \\
\psi^\downarrow_{1\over 2}(x, { k}_{\perp} )  &=&  \left( {k^1 - \iota k^2 \over x M_n} \right) \, \varphi^2_q(x,\, { k}_{\perp}) \,, \nonumber \\
\psi^\downarrow_{-{1\over 2}} (x, { k}_{\perp} ) 
&=&  \varphi^1_q(x,\, { k}_{\perp}) \,.
\label{wave}
\eea
The wavefunction $\varphi_q^i(x, { k}_{\perp})$ is the generalization of the  LFWFs derived from recent work on soft-wall holographic model in AdS/QCD \cite{stan2012}.  For the massless constituents the LFWF have the simple form
\be {\varphi^i_q(x,\, { k}_{\perp}) } =
 {4 \pi \over \kappa } \,  N^i_q\, \sqrt{\log(1/x) \over 1-x }\,
x^{a^i_{q}} {(1-x)}^{b^i_{q}} 
\, e^{ {-{k^2_{\perp} \over 2 \kappa^2\,} { { \rm log(1/x)} \over (1-x)^2 }
} }  \,. \label{varphi} \ee
Here $N^i_q$ is the normalization constant, $a^i_q$ and $b^i_q$ are the free parameters to be fitted to the experimental data on electromagnetic form factors, magnetic moments, and charge radii. It is important to mention here that the analytical form of frame independent wavefunction successfully predict the pion coupling constant, charge radius, space and time like behaviour of form factors \cite{valery2015}.

The expressions for GPDs for up and down quark in the LFQM are given as
\be
 H^q(x, q^2) = \frac{ n_q}{ I^q_1} 
 ({N^1_q})^2  \, x^{2a^1_{q} } {(1-x)}^{ 2 b^1_{q}+1} 
\left [ 1 + \sigma^2(x) \kappa^2    \left ( {1 \over  \log(1/ x)} -  {q^2_{\perp} \over 4 \kappa^2 } \right) \right] e^{- { \log(1/x) \over  4\kappa^2 } q^2_{\perp} } \,, 
\ee 
\bea
 E^q(x, q^2) &=& \frac{ 2\, \kappa_q}{ I^q_2}\,  N_q^1N_q^2  \, x^{a^1_{q} + a^2_{q} -1 }
  {(1-x)}^{ b^1_{q}+ b^2_{q}+2} e^{-  q^{2}_{\perp}  \log(1/x)  \over  4 \kappa^2 }\,,
\eea
 where $n_q$ is the number of valence quarks in the nucleon and $\kappa_q$ is the quark anomalous  magnetic moment.  The integrals in the above equation are defined as 
 \bea I_1^q  &=& \int^1_0 {\mathrm d} x\, 
(N^1_q)^2  \, x^{2a^1_{q} } {(1-x)}^{ 2 b^1_{q}+1} 
\left [ 1 + { \sigma^2(x) \kappa^2 \over \log(1 / x)} \right] \,, \\  
I^q_2 &=&  2  \int^1_0 {\mathrm d} x\, 
 N_q^1N_q^2  \, x^{a_{1q} + a_{2q} -1 } {(1-x)}^{ b_{1q}+ b_{2q}+2} \,,
\eea
with the convention $ \sigma(x) =  \frac{N^2_q}{N^1_q}  \,  x^{(a^2_{q} -a^1_{q}) } {(1-x)}^{  (b^2_{q} - b^1_{q} )}.$


\section{DGLAP evolution for Generalized Parton Distributions}
\label{dglap}
 
In this section we will discuss the inclusion of perturbative evolution effects in the GPDs calculated in the light front quark model.  The evolution of the GPDs with scale parameter is governed by the DGLAP-like equations, however, the integro-differential nature of equation makes its difficult to find analytical solutions \cite{dglap1,dglap2,dglap3,dglap4}. In literature there exist numerous techniques, such as, the so-called brute-force method \cite{brute}, Laguerre polynomials \cite{laguerre}, Mellin moment space transformation with subsequent inversion \cite{melin},  QCDFIT program \cite{qcdfit}, etc.. Following the work of Ref. \cite{majid1,majid2} where a numerical procedure is used to obtain evolved GPDs in the AdS/QCD approach, we will use the same numerical technique to evolve the GPDs in LFQM.  


The independence of physical observables from  scale parameter $\mu$, gives the following type of DGLAP like equation for valence quark GPD $H^q (x, t)$:
\bea
\mu^2 { {\rm d} \over  {\rm d} \mu^2} H^q (x, t, \mu^2)  &=&  \left ({\alpha_s  \over 2 \pi}  \right)\int_x^1 { {\rm d} z \over z }
 \left[ P \left( {x \over z} \right) \right]_+ H^q (z, t, \mu^2)\,, 
 \label{dg}
 \eea
 where $[....]_+$ is the usual ``plus regularization'' scheme for the  DGLAP evolution kernel  \cite{melin}.  The leading order quark-quark splitting function $P(z) = C_f ({1 + z^2 \over 1-z}) $ with $C_f = {N^2-1 \over 2 N}$ gives the probability of a quark after being  radiating a gluon is left with momentum fraction $z$  of the original momentum. The term corresponding to the gluon splitting function is not considered in this prescription as we are considering  only the valence quarks contribution in the calculations of GPDs.  It is also worth to mention here that DGLAP equation perfectly works for ordinary parton distributions with $t=0$.

The basic idea in solving the Eq. (\ref{dg}) is to absorb the uncalculable perturbative effects into the modified GPDs also called as evolved GPDs. We can rewrite the Eq. (\ref{dg}) as
\bea
H^q (x, t, \mu^2)  &=& H^q (x, t, \mu_0^2)  + \left ({\alpha_s  \over 2 \pi}  \right) 
\left ( \ln { \mu^2 \over \mu_0^2 }\right) \int_x^1 { {\rm d} z \over z } \, P \left( {x \over z} \right)_+ H^q (z, t, \mu_0^2) + {\cal O} (\alpha_s^2)\,. 
 \label{dg1} \eea
 The convolution integral on the right hand  side of  Eq. (\ref{dg1}) can be easily simplified as\be  \int_x^1 {{\mathrm d} z \over z} P \left( \frac{x}{z} \right)_+ H^q(z,t ) =  \int_x^1 {{\mathrm d} z \over z}  P \left(\frac{x}{z} \right)  \left( H^q(z, t) - \frac{x}{z} H^q(x, t ) \right)
-  H^q(x, t) \int_0^x {\mathrm d} z P(z)\,. \ee
Further, we need the physical coupling constant at different energies i.e. the running coupling constant  $\alpha_s $ expressed as a function of renormalization scale  $\mu$. In leading order approximation, running coupling constant  $\frac{\alpha_s (\mu^2) }{2 \pi} = \frac{2}{ \beta_0 \,  {\mathrm ln} (\mu^2 / \Lambda^2)}\,,  $  where $\beta_0=11-2/3\, n_f$  and $\Lambda$ is the QCD scale parameter  \cite{pdg}. Using this prescription, we have  solved the  Eq.(\ref{dg}) numerically for different  values of $x$ for the initial value chosen as  $\mu_0=0.3$ GeV.  It is also important to mention that the evolution equation for the GPDs $E^q(x, t)$ is same as $H^q(x, t)$.  

\begin{figure}[htbt]
\begin{minipage}[c]{0.98\textwidth}
\small{(a)}
\includegraphics[width=7cm,height=5cm,clip]{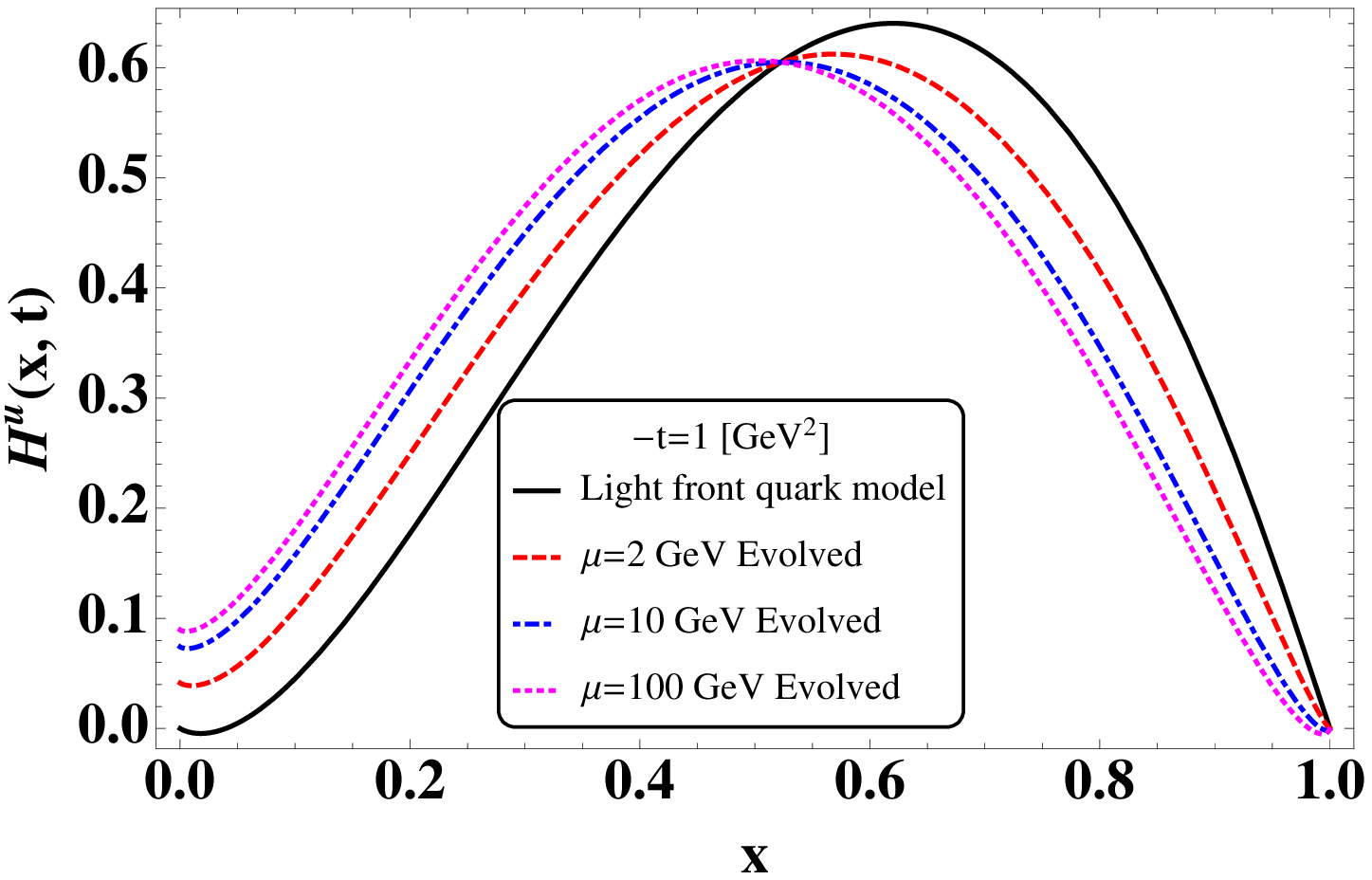}
\small{(b)}\includegraphics[width=7cm,height=5cm,clip]{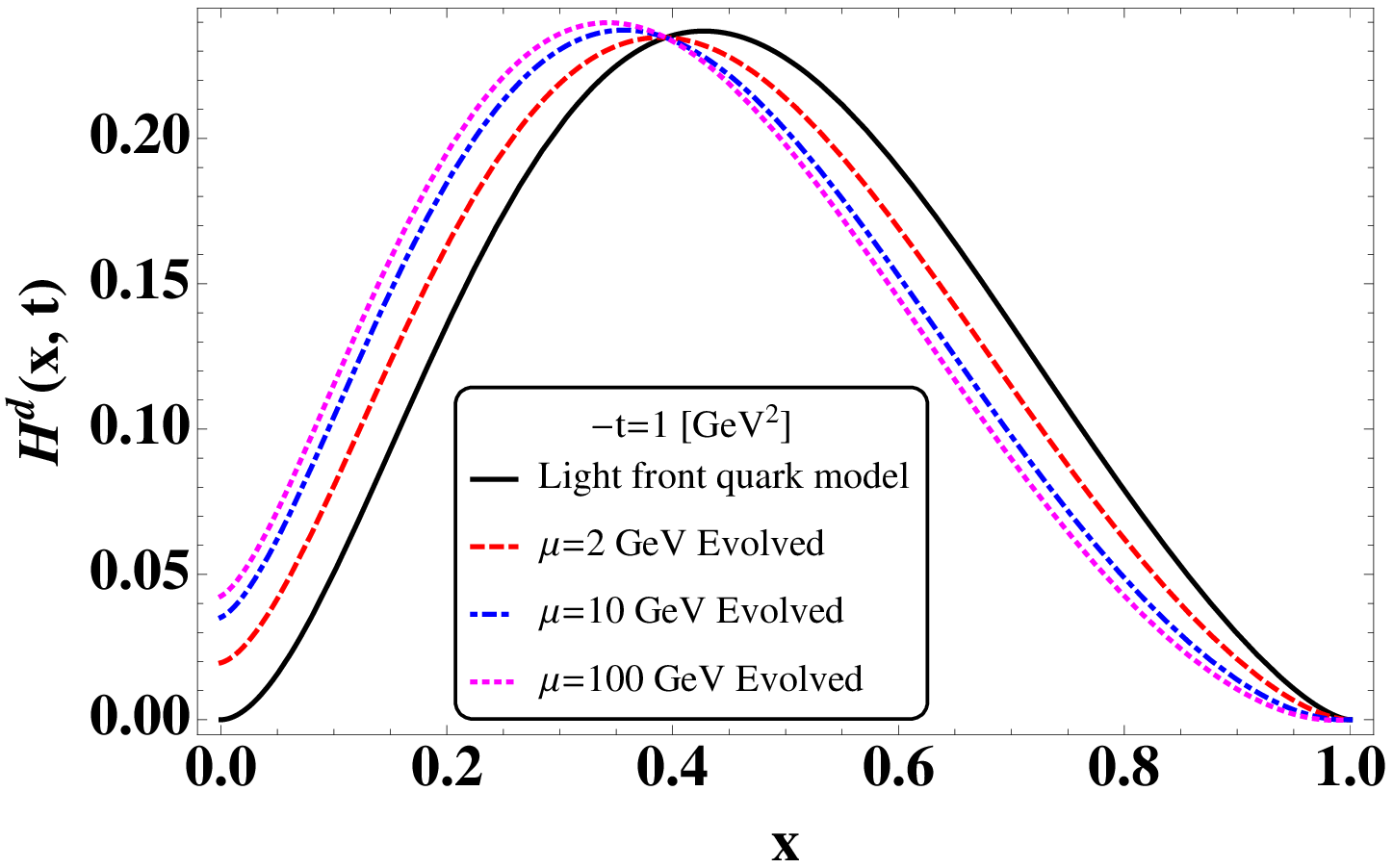}
\end{minipage}
\begin{minipage}[c]{0.98\textwidth}
\small{(c)}
\includegraphics[width=7cm,height=5cm,clip]{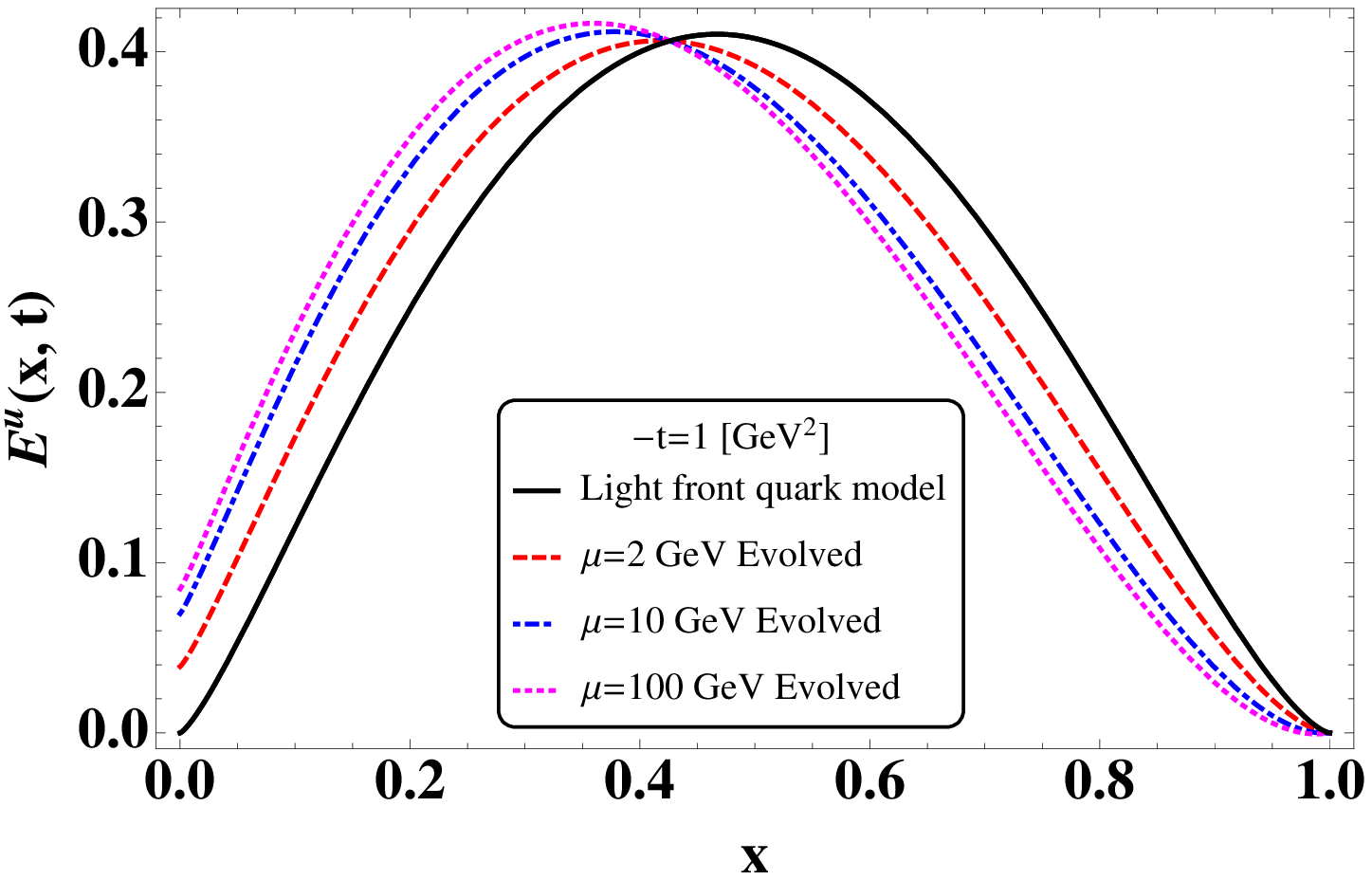}
\hspace{0.1cm}
\small{(d)}
\includegraphics[width=7cm,height=5cm,clip]{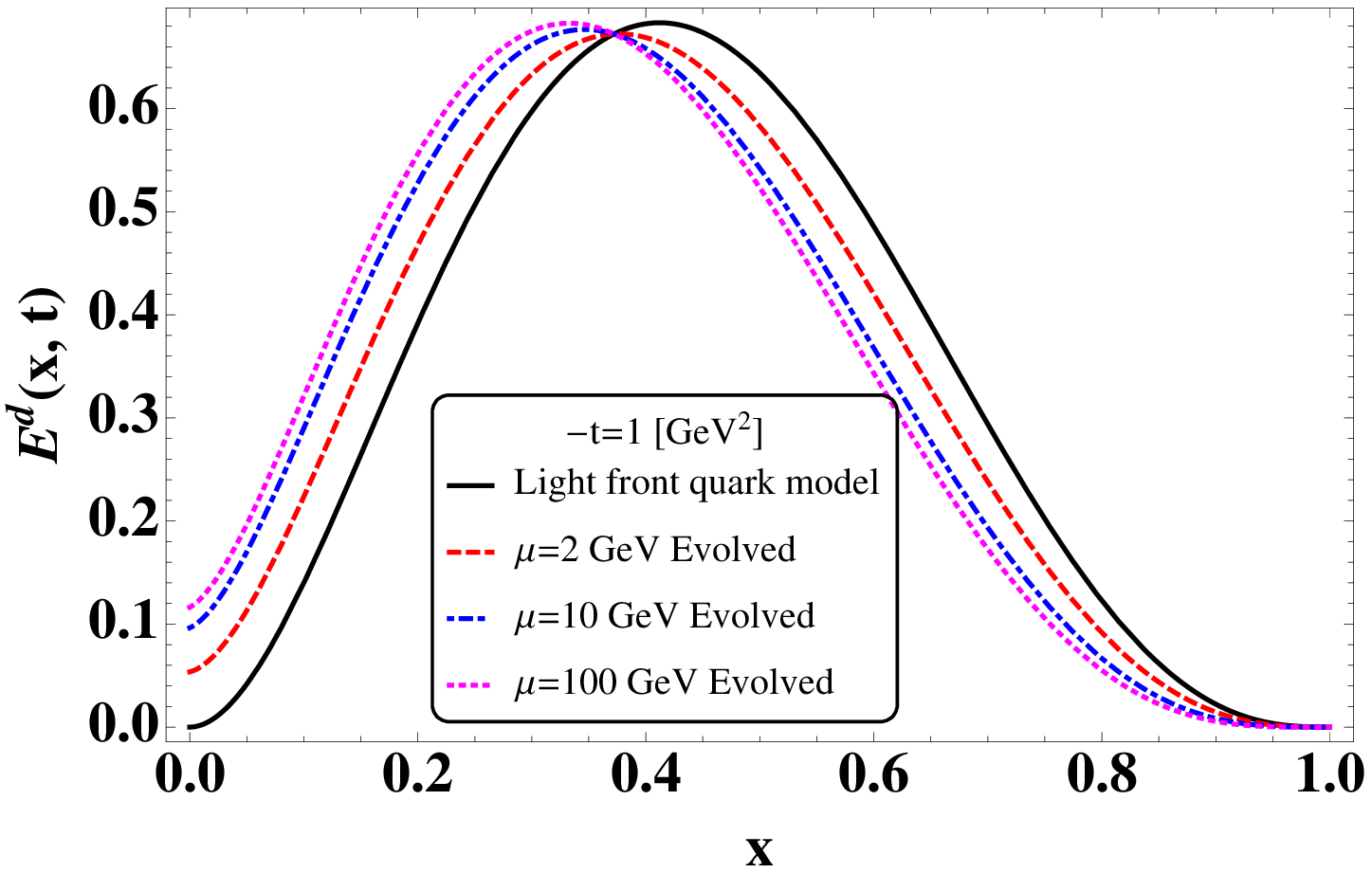}
\end{minipage}
\caption{
\label{hu}(Color online). Plots of (a) the evolved generalized parton distributions $H^u(x,t,\mu)$ vs $x$ for fixed values of $ -t =1$ GeV$^2$ and $\mu =2,10,100$ GeV for $u$ quark, (b) for $d$ quark, (c) $E^u(x,t,\mu)$ vs $x$ for fixed values of $ -t$ and $\mu$ for $u$ quark, and (d) for $d$ quark.}
\end{figure}

We will now discuss the  behavior of evolved  GPDs $H^q(x,t,\mu)$  and $E^q(x,t,\mu)$  with $x$ for the various values of the scale parameter. In Fig. \ref{hu}(a) and (d), we have presented the evolved GPDs $H^q(x,t,\mu)$ and   $E^q(x,t,\mu)$ as a function of $x$ for fixed values of $-t = 1$ GeV$^2$  for the up and down quark.  In order to understand the implication of the  DGLAP evolution on both GPDs, we have presented the results with $\mu =2,10,100$ GeV and also presented results in LFQM.   It is clear that the qualitative behaviour of  both GPDs with parameter $x$ is same for up and down quark.  Both the GPD $H(x, t)$ and $E(x, t)$ increase with $x$, obtain a maxima and then falls to zero as $x \to 1$. For the evolved GPDs, peak shift towards a lower value of $x$ for the higher values of scale parameter $\mu$ as the probability of a gluon being radiated is higher at large values of $x$, hence the distribution shift towards lower value of $x$ for all the cases. We observe that the maxima for evolved GPDs for different $\mu$ remain same as  the LFQM results except for the $H^u(x,t,\mu)$.

\section{ GPDS in impact parameter space}
\label{impact}
We are interested in understanding the impact of evolution effects on the GPDs in the impact parameter space. For zero skewness, the momentum transfer is only in the transverse direction, thus the Fourier transform of GPDs with respect to the  momentum transfer gives the transverse distribution of the partons in impact parameter space \cite{impact}.  GPDs in the  transverse impact parameter space  give the probability  of finding a  parton in the transverse plane, which is an important aspect while studying the nucleon structure. Impact parameter space GPDs are defined as
\bea
q(x,\, b,\,\mu)  &=& {1 \over {(2 \pi) }^2} \int {\mathrm d}^2 q_{\perp} e^{-  b_{\perp}. q_{\perp } } H(x, \, t,\,\mu) \,,  \\
e^q(x,\, b,\,\mu) &=& {1 \over {(2 \pi) }^2} \int {\mathrm d}^2 
q_{\perp} e^{- b_{\perp}. q_{\perp } } E(x, \, t,\,\mu) \,.\eea
The transverse impact parameter $b =|b_{\perp}|$ is a measure of the transverse distance between the struck parton and the center of momentum of  the hadron and satisfies the condition $\sum_i x_i b_i =0$,  where the sum is over the number of partons.  Impact parameter GPDs give an estimate of the size of the bound state, however, in order to have an exact estimate of nuclear size, we need to understand the spatial extension as well.

\begin{figure}[htbp]
\begin{minipage}[c]{0.98\textwidth}
\small{(a)}
\includegraphics[width=7cm,height=5cm,clip]{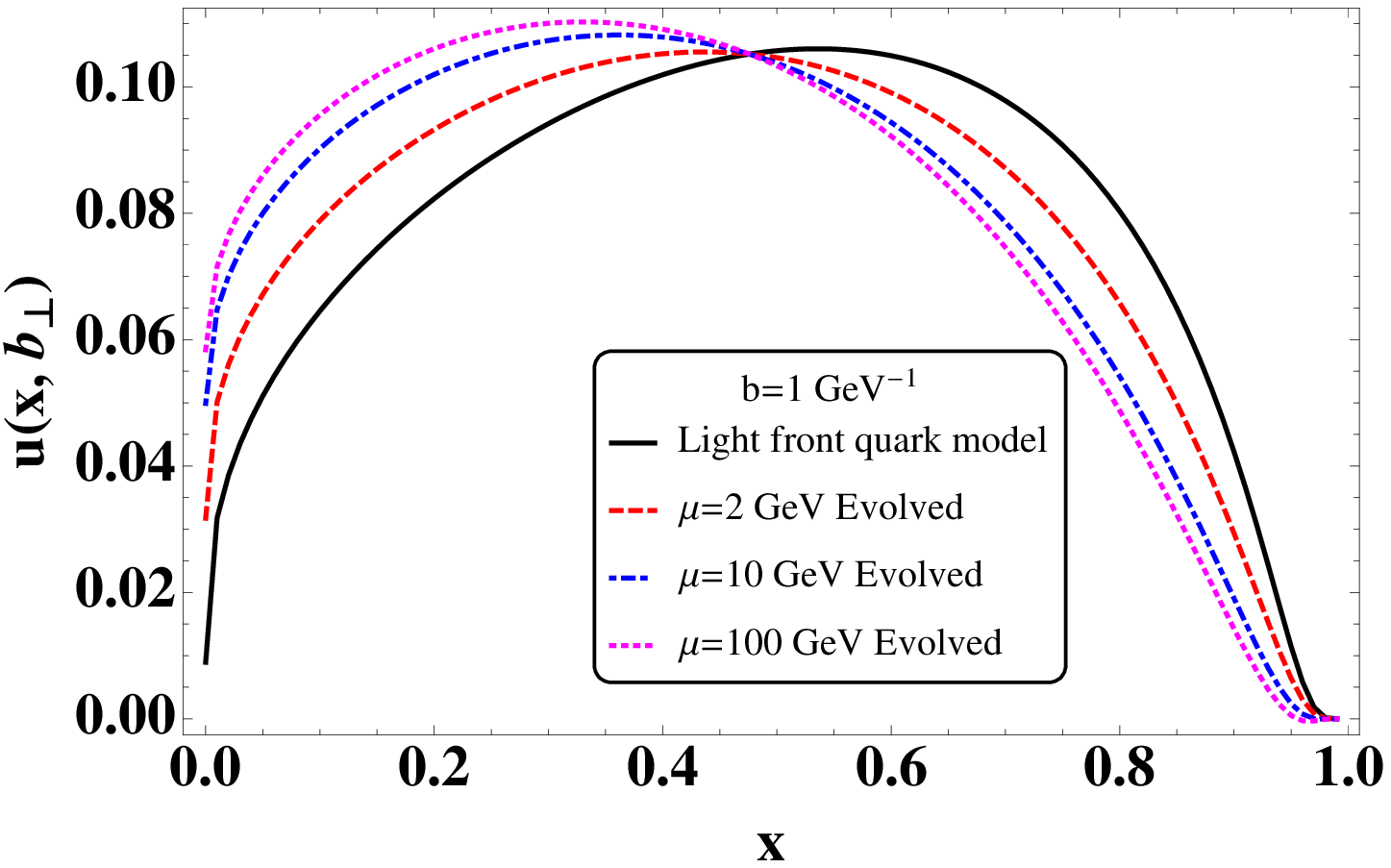}
\hspace{0.cm}
\small{(b)}
\includegraphics[width=7cm,height=5cm,clip]{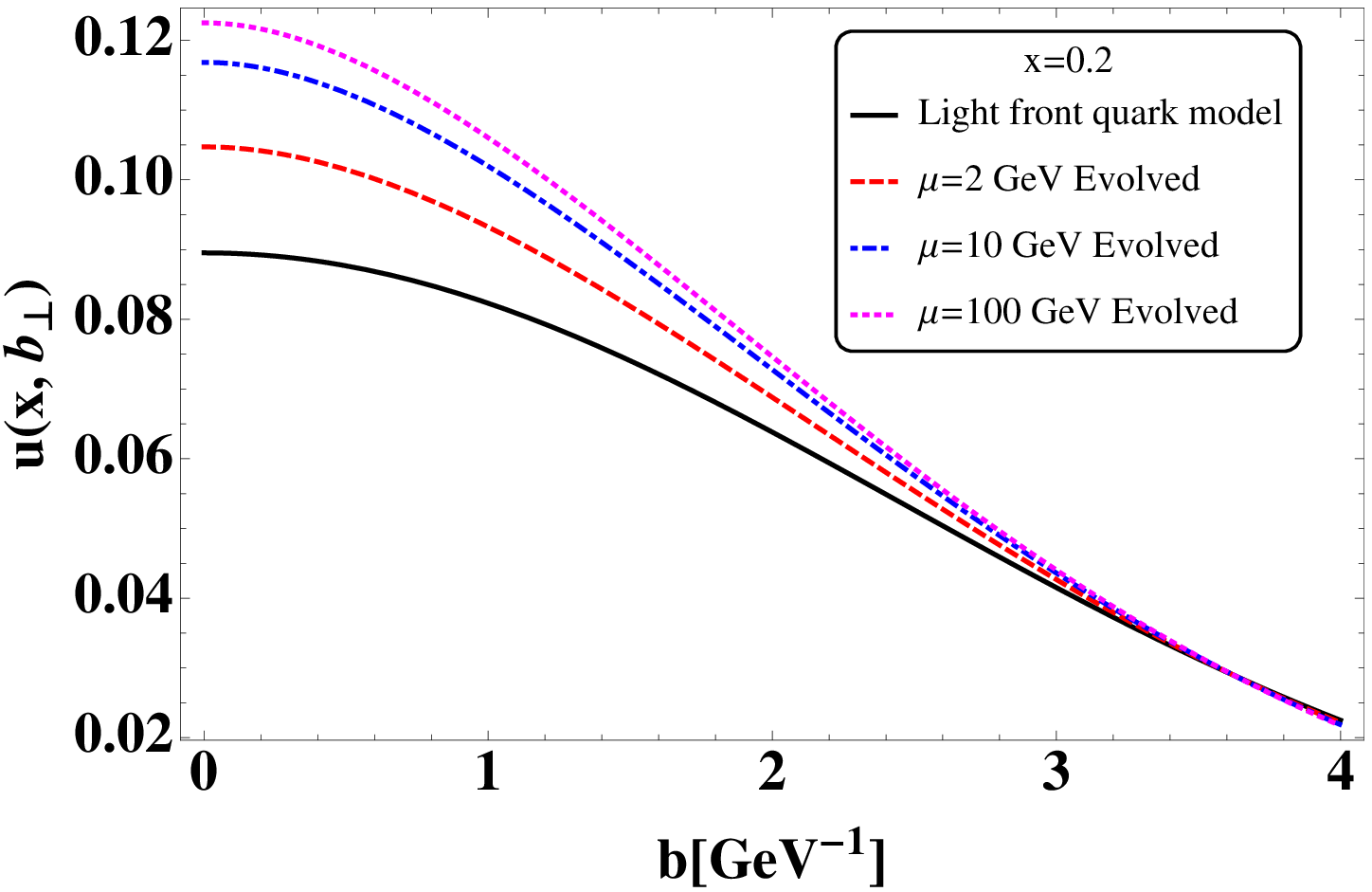}
\end{minipage}
\begin{minipage}[c]{0.98\textwidth}
\small{(c)}
\includegraphics[width=7cm,height=5cm,clip]{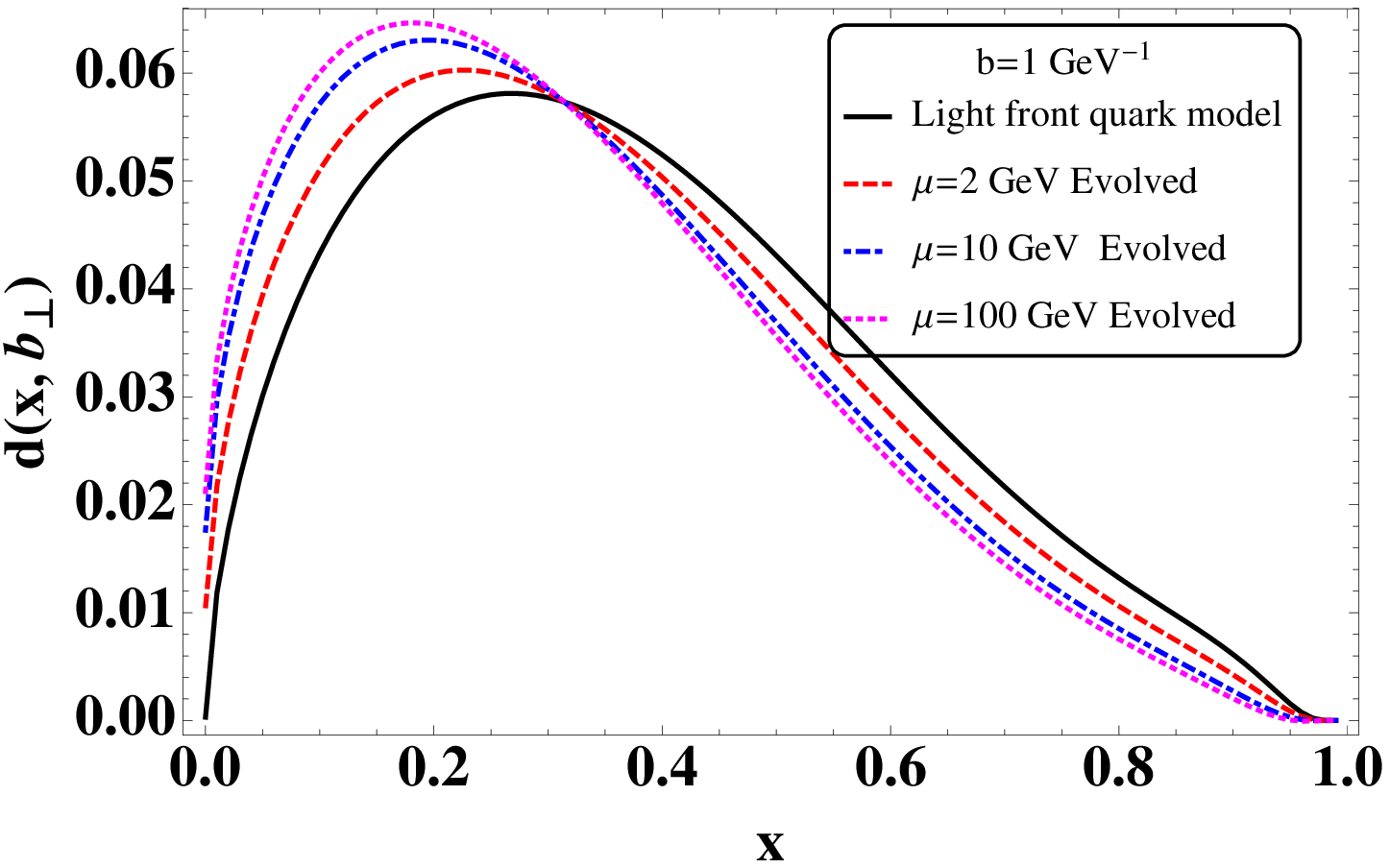}
\hspace{0.1cm}
\small{(d)}
\includegraphics[width=7cm,height=5cm,clip]{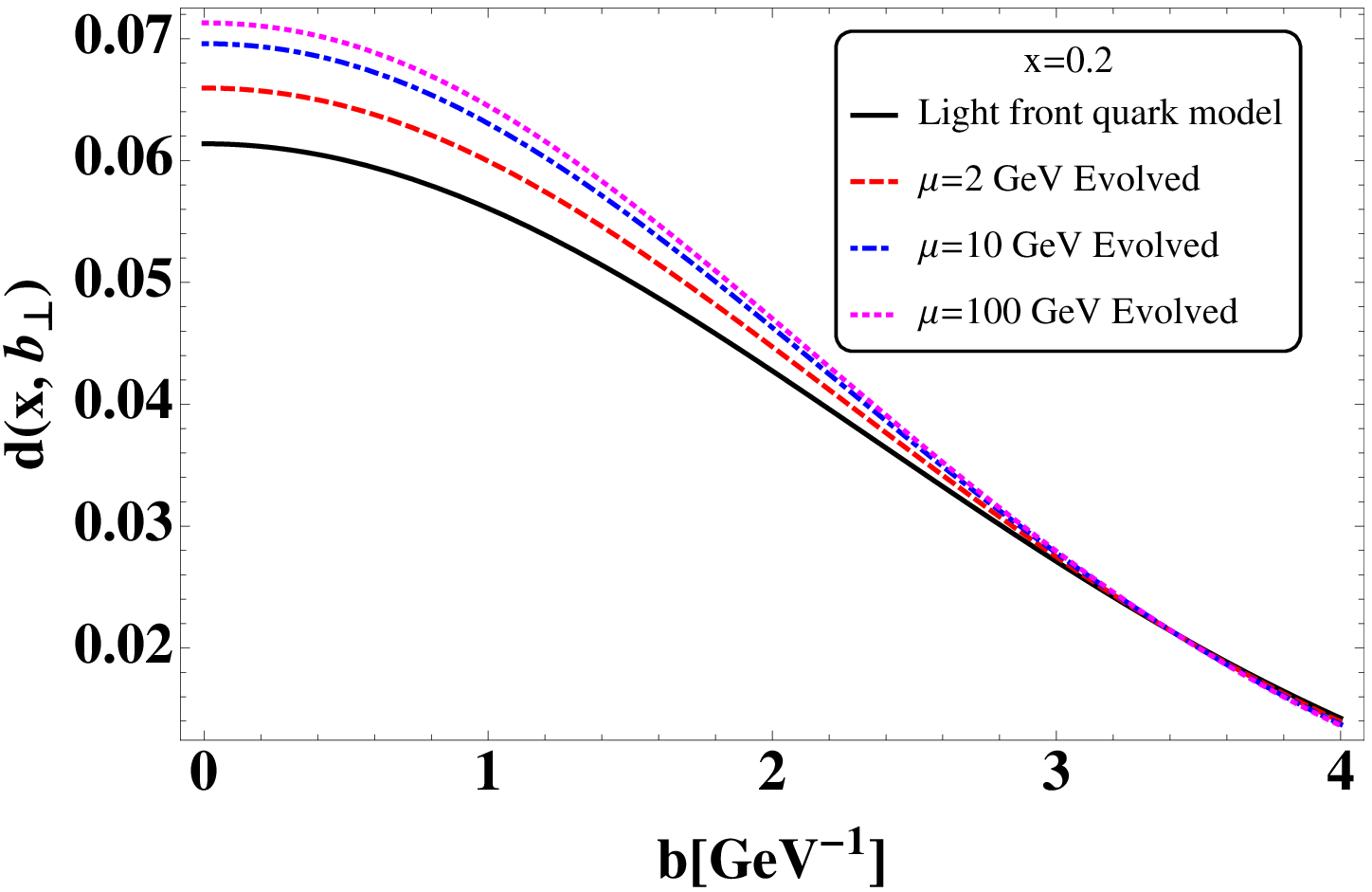}
\end{minipage}
\caption{\label{ihub}(Color online). Plots of (a) $u(x,b,\mu)$ vs $x$ for fixed values of $b = 1$ GeV$^{-1}$,  (b)  $u(x,b,\mu)$ vs the impact parameter $b = | b_\perp |$ for $x=0.2$   
(c) same as (a) but for $d$ quark, and (d) same as  (b) but for $d$ quark.}
\end{figure}

\begin{figure}[htbp]
\begin{minipage}[c]{0.98\textwidth}
\small{(a)}
\includegraphics[width=7cm,height=5cm,clip]{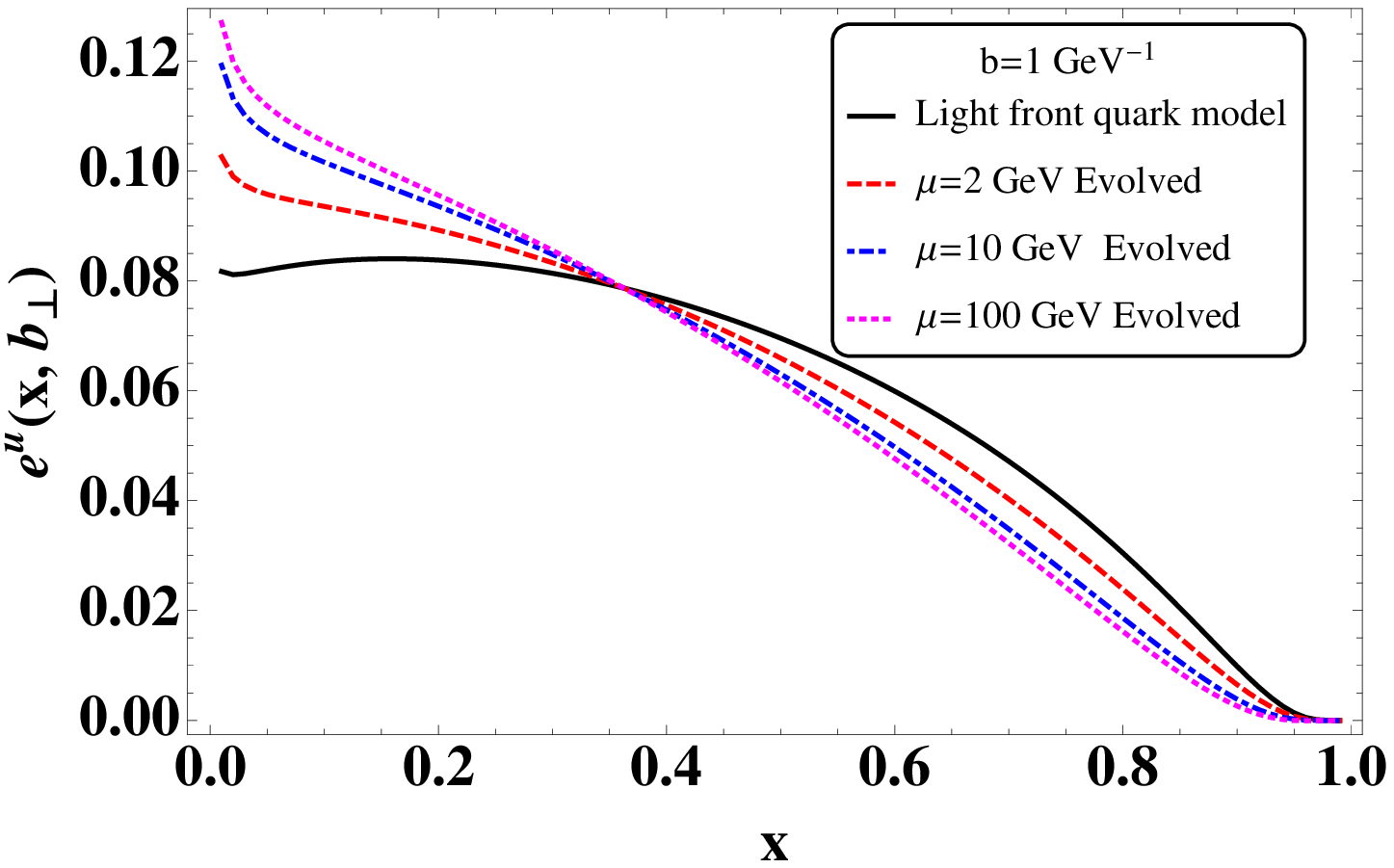}
\small{(b)}
\includegraphics[width=7cm,height=5cm,clip]{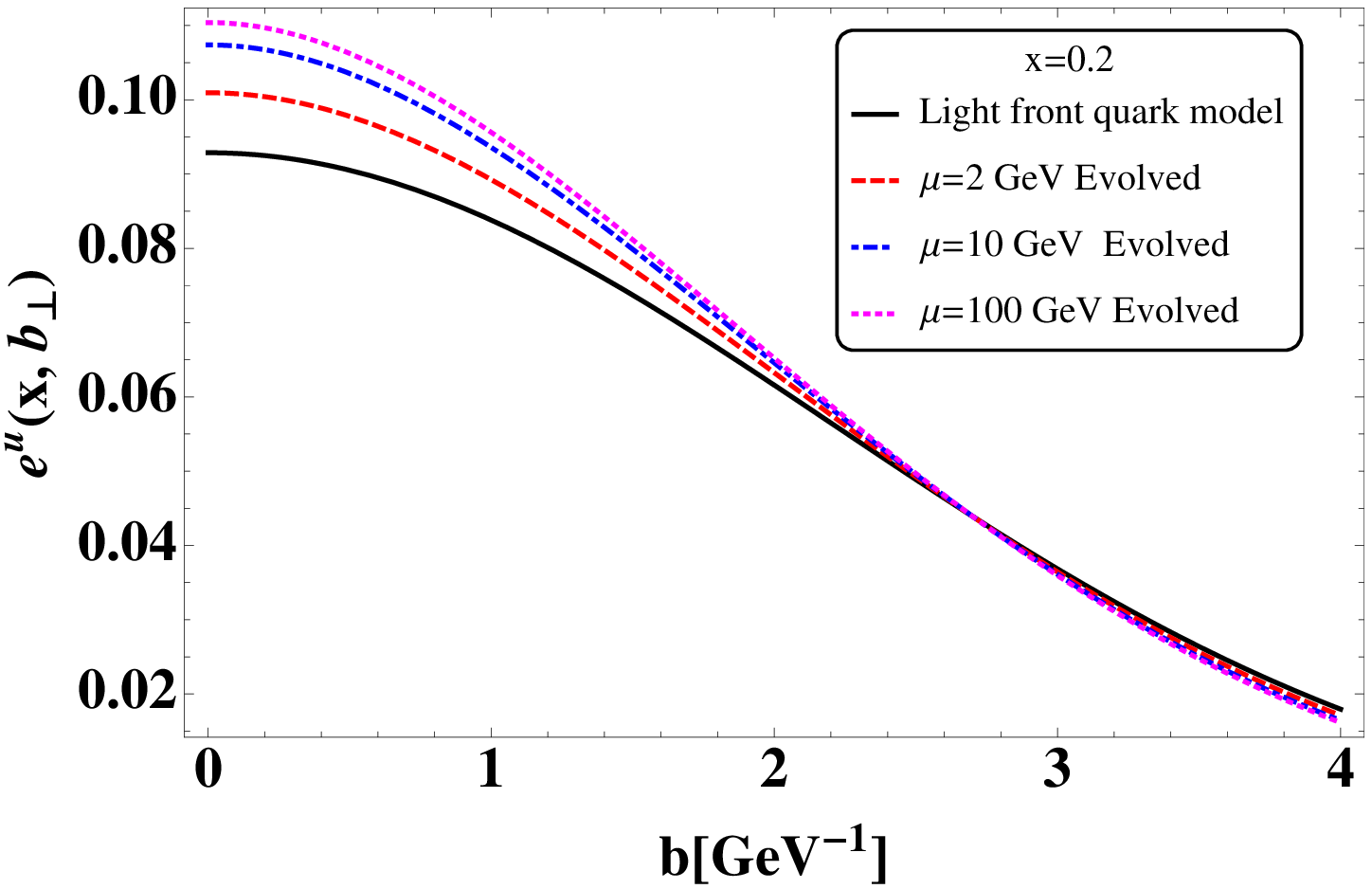}
\end{minipage}
\begin{minipage}[c]{0.98\textwidth}
\small{(c)}
\includegraphics[width=7cm,height=5cm,clip]{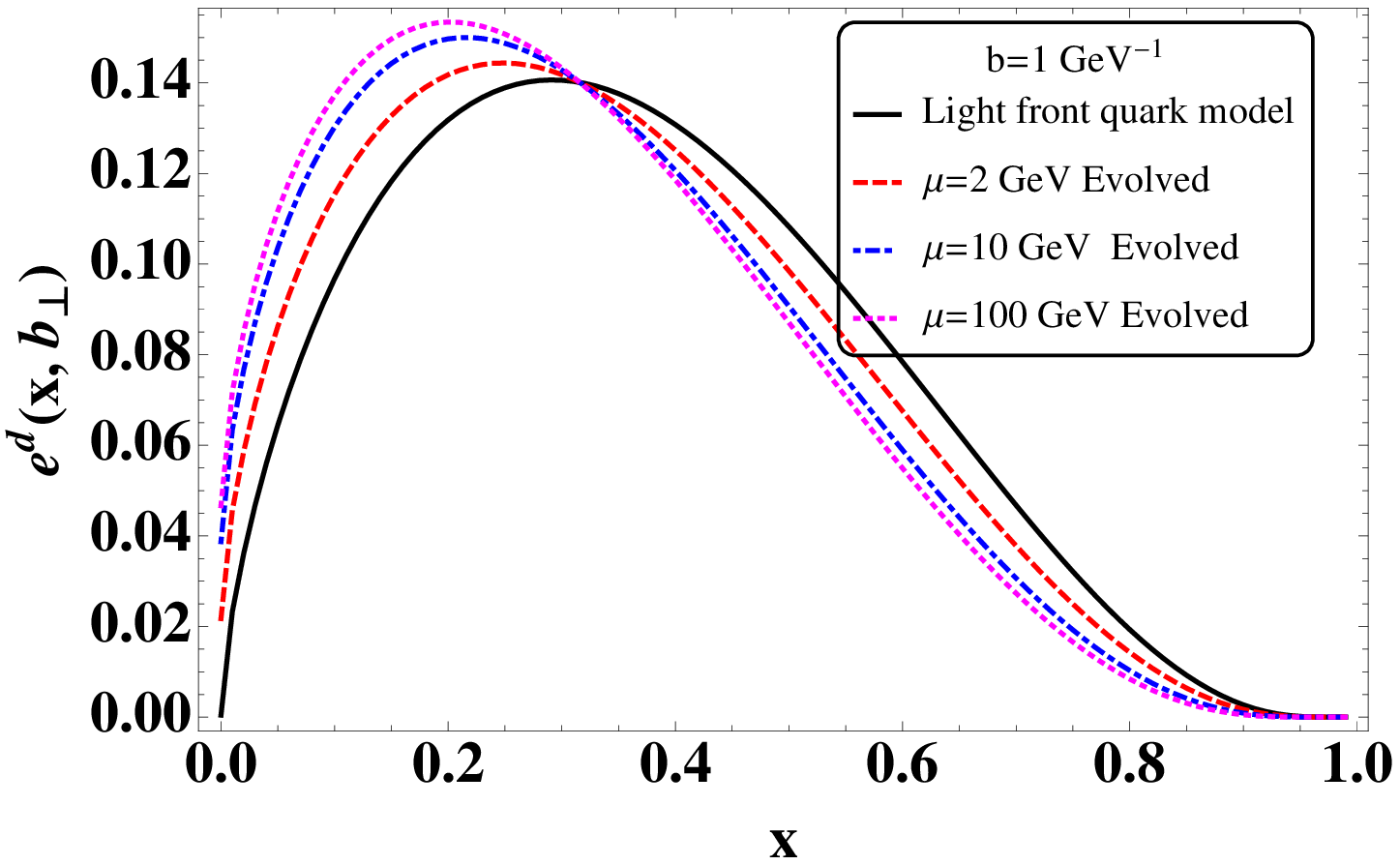}
\small{(d)}
\includegraphics[width=7cm,height=5cm,clip]{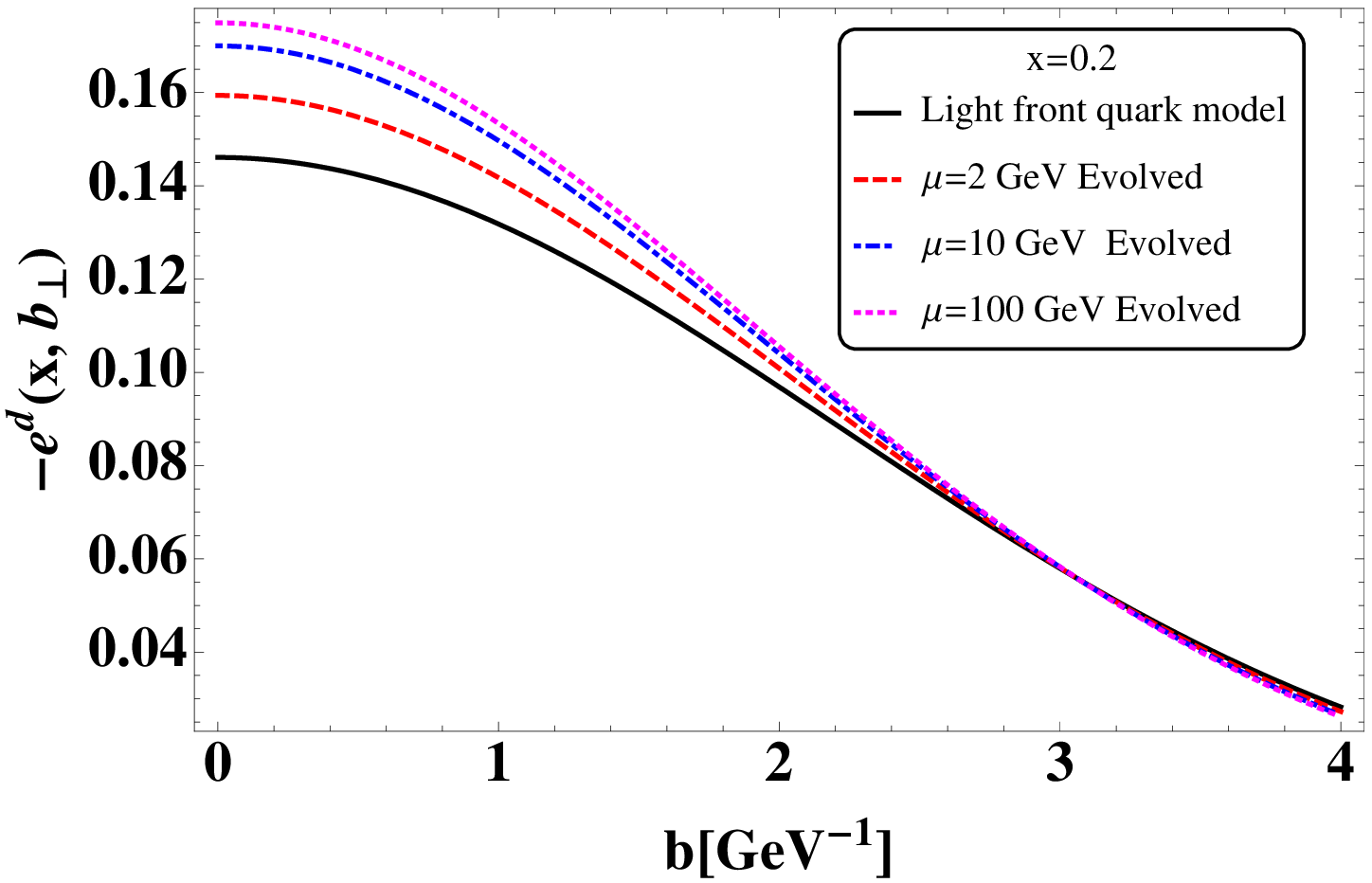}
\end{minipage}
\caption{\label{ieub}(Color online). Plots of (a) $e^u(x,b,\mu)$  vs $x$ for fixed values of $b = 1$ GeV$^{-1}$,  (b)  $e^u(x,b,\mu)$ vs the impact parameter $b = | b_\perp|$ for $x=0.2$   
(c) and (d) are same as in (a) and (b) but for $d$ quark.}
\end{figure}

Now we investigate the implications of evolution on the impact parameter dependent GPDs. Since  the impact parameter GPDs are a function of three variable $x,b,\mu$, we consider the variation of GPDs with $x $ and $b$ separately, for different values of $\mu$. In Fig. \ref{ihub}(a), we have plotted the behaviour of evolved GPDs $u(x,b,\mu)$ with $x$ for fixed values of $b=1$ GeV$^{-1}$ and in Fig. \ref{ihub}(b) we have plotted the behaviour of same GPD with the impact parameter $b$ for the fixed values $x =0.2$. In Figs. \ref{ihub}(c) and \ref{ihub}(d), we plot the same GPDs for the down quark using the same set of parameters. In order to understand the implications of the scale parameter, we have used the different set of scale parameter $\mu=2,10,100$ GeV.  Similar plots showing the behaviour of  GPDs $e^q (x, b,\mu)$ for both up and down quarks are shown in Fig. \ref{ieub}.

One can observe that the effect of evolution is more prominent in the impact parameter space than  momentum space. The maxima of GPDs increases significantly as $\mu$  increases,  which implied that the magnitude of GPDs is maximum at the centre of nucleon and increases further for the larger $\mu$. Both the GPDs $q(x,b,\mu)$ and $e^q(x,b,\mu)$  in the evolved approach and LFQM converges for the higher values of $x$ and portray a similar behaviour for up and down quark.   It is also interesting to observe that for the small values of  $b$, the magnitude of GPD $q(x, b,\mu)$ is larger for up quark than down quark, whereas the magnitude of the GPD $e^q(x,b,\mu)$ is marginally larger for down quark than  up quark.   Further, we observe that in both cases, the maxima of GPDs shifted towards a lower value of $x$ as $b$ increases, therefore the  transverse profile is peaked at $b=0$ and falls off further. 

We also wanted to make a point that the  hadronic form factors $F^{1,2} (t)$ are independent of evolution as they are related to the GPDs when the $x$ dependence in being integrated out. Since the behavior of GPDs are not very well established experimentally and there are only phenomenological methods, future experimental  information on the GPDs could  render the  present situation more precise.

\section{Summary and conclusion}
\label{conc}
We investigated the effect of perturbative evolution on GPDs for up and down quarks in nucleon using the effective light-front quark model. We  compared our results for the evolved GPDs  the momentum space with LFQM results.   A detailed comparison of behaviour of evolved GPDs with the LFQM in  impact parameter space is also presented as the impact of evolution is more significant  in the transverse impact parameter space.  We have shown the effect of evolution on the GPDs in the different regions of kinematics. It is observed that the magnitude of GPDs increases at the center of nucleon   for the large values of evolution parameter.  In future, we plan to generalize the LFWFs to sea quarks, antiquarks, and gluons, which could then be used in the evaluation of different hadronic processes so that one can directly compare the theoretical results with the experiments.



\section*{Acknowledgements}
 N.S.  would like to thank  Majid Dehghani  for useful discussions.  This work is supported by Department of Science and Technology, Government of India (Grant No. SR/FTP/PS-057/2012).

\end{document}